\documentclass[]{interact}

\usepackage{epstopdf}
\usepackage[caption=false]{subfig}

\usepackage[numbers,sort&compress,merge]{natbib}
\bibpunct[, ]{[}{]}{,}{n}{,}{,}

\theoremstyle{plain}

\usepackage{graphicx}

\usepackage[american]{babel}
\begin{document}

\title{First principles absorption spectra of Au nanoparticles: from quantum
to classical}

\author{
\name{
Samuel Hernandez\textsuperscript{a}$^\ast$\thanks{$^\ast$These authors contributed equally to this work.}, 
Yantao Xia\textsuperscript{b}$^\ast$, 
Vojt\v{e}ch Vl\v{c}ek\textsuperscript{a}, 
Robert Boutelle\textsuperscript{a}, 
Roi Baer\textsuperscript{c},  
Eran Rabani\textsuperscript{d,e},  
Daniel Neuhauser\textsuperscript{a}}
\affil{
\textsuperscript{a} Department of Chemistry and Biochemistry, University of California, Los Angeles, California 90095, USA.; 
\textsuperscript{b} Department of Chemical and Biomolecular Engineering, University of California, Los Angeles, California 90095, USA.;
\textsuperscript{c} Fritz Haber Center for Molecular Dynamics and Institute of Chemistry, The Hebrew University of Jerusalem, Jerusalem 9190401, Israel;
\textsuperscript{d} Department of Chemistry, University of California, Berkeley, California 94720, USA.
\textsuperscript{e} The Raymond and Beverly Sackler Center for Computational Molecular
and Materials Science, Tel Aviv University, Tel Aviv 69978, Israel.
}
}
\maketitle
\begin{abstract}
Absorption cross-section spectra for gold nanoparticles were calculated
using fully quantum Stochastic Density Functional Theory and a classical
Finite-Difference Time Domain (FDTD) Maxwell solver. Spectral shifts
were monitored as a function of size (1.3\textendash 3.1~nm) and
shape (octahedron, cubeoctahedron, and truncated cube). Even though
the classical approach is forced to fit the quantum TDDFT at 3.1nm,
at smaller sizes there is a significant deviation as the classical
theory is unable to account for peak splitting and spectral blue shifts
even after quantum spectral corrections. We attribute the failure
of classical methods at predicting these features to quantum effects
and low density of states in small nanoparticles. Classically, plasmon
resonances are modeled as collective conduction electron excitations,
but at small nanoparticle size these excitations transition to few
or even individual conductive electron excitations, as indicated by
our results. 
\end{abstract}
\maketitle

\section*{Introduction}

The unique physical and chemical properties of nanoparticles have
generated intense academic and industrial interest, in hope that these
properties, once well-understood, could be used for technological
advances. nanoparticle materials exhibit physical and chemical properties
very different from those of their bulk counterparts, often resulting
from enhanced surface interactions or quantum effects \cite{BulktoNano}.
For example, noble metal nanoparticles are drawing intense interest
because of their ability to sustain localized surface-plasmon resonances
(LSPRs) \cite{LSPRReview}. LSPRs are collective oscillations of surface
conduction band electrons excited by an oscillating electric field,
typically a photon. These oscillations enable strong absorption and
scattering of subwavelength structures. Coupling of photons to conduction
band electrons at metal interfaces improves efficiency for ultrafine
sensing methods \cite{LSPRsensors,LSPRsensors2}, enhanced catalysis
\cite{cat}, energy-transfer \cite{ETran1,ETran2}, and has enabled
applications such as light concentrators in solar cells \cite{solcon}
and cancer therapies \cite{cancer}.

Experimental characterization of metallic nanoparticles is extremely
difficult, since optical detection in the far-field is hampered by
low signal to noise ratio due to low scattering and absorption intensities
\cite{huffman}. The LSPR peaks are further broadened and damped as
the size of the particle decreases below the electron mean free path
(40 nm for gold \cite{EMeanFreePath}). Finally, the spectral properties
are strongly coupled to the stoichiometry, size, shape, and surrounding
medium \cite{LSPRReview,plasshape1,plasshape2} making an \textit{a
priori} prediction difficult. The difficulty in observing broad, weak
LSPR signals has led to conflicting results from experiments involving
quantum-sized plasmonic particles, with multiple reports of LSPR redshifts
or blueshifts as the size is reduced \cite{plasshift1,plasshift2}.

Theoretical investigations often rely on classical methods, such as
Mie Theory \cite{Mie} and Finite-Difference Time-Domain (FDTD) Maxwell
Solvers \cite{MEEP} which however inherently do not capture quantum
effects important at scales of few nm. Size-dependent electron scattering
terms have been included in the classical approaches, but the use
of these correction results in predictions with redshifts and imperceptible
plasmon resonances at small size scales \cite{imperceptible} that
conflict with experimental findings \cite{DOSAg,finding1,finding2}.
Additionally, the classical models assume a density of states (DOS)
sufficiently populated in the Fermi level such that the LSPRs are
a collective electron oscillation. This picture is however challenged
by experiments, which suggest that small clusters (below 3nm) exhibit
nonmetallic character so that with decreasing size, discrete peaks
appear in their optical spectra \cite{DOSAu,DOSAg}.

In this work fully first-principles quantum methods such as time-dependent
density functional theory (TDDFT) are used to investigate the transition
to the quantum regime of photo absorption cross sections in gold nanoparticles.
Observation of the transition in gold requires applying TD-DFT to
large nanoparticles including several hundreds of atoms. For this,
we employ the recently developed stochastic approach to electronic
structure \cite{sDFT,Neuhauser2014,Rabani2015} and in particular
the stochastic TDDFT (sTDDFT) approach \cite{dan_jcp}, which allows
linear-scaling effort with respect to system size. We studied a range
of stable \cite{barnard_1,barnard_2} closed-shell nanoparticles containing
between 44 and 344 atoms, corresponding to diameters of 1.34 to 3.12
nm. By comparing with classical Maxwell simulations, we find that
systems having less than $\sim$200 atoms exhibit strong quantum signatures
(appearance of new absorption maxima and peak splitting) which depend
on shape of the nanoparticle and are missing in the FDTD results.

\section*{Theory}

In this section, we begin by reviewing the theory of absorption of
light by small particles, in both the quantum-mechanical and classical
picture. Next, we present the respective implementation of the quantum
and classical theories in computational chemistry, namely the stochastic
TDDFT and the FDTD Maxwell Equations, and explain their merits and
limitations. 

Within the linear response approach, the photo absorption cross section
is given by 
\begin{equation}
\sigma(\omega)=\frac{e^{2}}{3\epsilon_{0}c}\omega\int\mathbf{r}\cdot{\chi}(\mathbf{r},\mathbf{r}^{\prime},\omega)\cdot\mathbf{r^{\prime}}\,d\mathbf{r}d\mathbf{r}^{\prime},\label{sigma_1}
\end{equation}
where polarizability relates induced charge density $\delta n$ and
external perturbing potential $\delta v$: 
\begin{equation}
\chi\left({\bf r},{\bf r}^{\prime},t-t^{\prime}\right)=\frac{\delta n\left({\bf r},t\right)}{\delta v\left({\bf r}^{\prime},t^{\prime}\right)}.\label{chi_lin_resp}
\end{equation}

The external perturbing potential takes form of a dipole that perturbs
the system instantaneously at $t=0$: 
\begin{equation}
\delta v\left({\bf r},t\right)=\gamma r_{i}\delta\left(t\right)\label{vpert}
\end{equation}
where $r_{i}=x,y$ or $z$ is one of the components of the Cartesian
vector ${\bf {r}}$. Here, the particles we consider are symmetric
so the absorption spectrum will be identical for all polarization
directions. Therefore, with no loss of generality $i=1$ and $r_{_{i}}=x$
below. 

The impulsive perturbation excites the system at all frequencies.
The perturbed system is propagated in time, and the induced dipole
moment signal is computed:
\begin{equation}
\mu_{i}(t)=\int r_{i}\delta n(\mathbf{r},t)d{\bf r}.\label{dipole_correlation}
\end{equation}
Finally the absorption cross section $\sigma\left(\omega\right)=\sum_{i=x,y,z}\sigma_{ii}\left(\omega\right)$
is obtained from: 
\begin{align}
\sigma_{ii}(\omega) & =\frac{e^{2}}{\epsilon_{0}c}\omega\int_{0}^{\infty}dte^{i\omega t}\mu_{i}\left(t\right).\label{dft:spectrum}
\end{align}

We compute the time-evolution of the induced charge density $\delta n\left(\mathbf{r},t\right)$
using first-principles time-dependent DFT, starting from the ground
state Kohn-Sham (KS) system with Hamiltonian (in atomic units) 
\begin{equation}
H\left[n\right]=-\frac{1}{2}\nabla^{2}+v_{ext}\left(\mathbf{r}\right)+v_{H}\left[n\right]\left(\mathbf{r}\right)+v_{xc}\left(n\left(\mathbf{r}\right)\right),\label{KSHamiltonian}
\end{equation}
where $v_{ext}\left(\mathbf{r}\right)$ is the external (nuclear)
potential energy and $v_{H}\left[n\right]\left(\mathbf{r}\right)=\int n\left(\mathbf{r'}\right)\left|\mathbf{r}-\mathbf{r'}\right|^{-1}d\mathbf{r'}$
is the Hartree potential. The last term $v_{xc}\left(n\right)$ is
the exchange-correlation potential in the local density approximation
(LDA)\cite{Perdew1992}.

The Hamiltonian is associated with a complete set of eigenstates $\left\{ \phi_{j}\left(\mathbf{r}\right)\right\} $
and corresponding eigenvalues $\left\{ \varepsilon_{j}\right\} $
($j=1,2,...$ is the state index) which are used as initial states
of the system at time $t=0$ when the perturbation of Eq.~(\ref{vpert})
is applied. At this moment and for subsequent times $t\ge0$ the total
density is a weighted sum of the instantaneous state densities $\left|\phi_{j}\left({\bf r},t\right)\right|^{2}$
\begin{equation}
n\left(\mathbf{r},t\right)=2\sum_{j}f_{\beta}\left(\varepsilon_{j},\mu\right)\left|\phi_{j}\left(\mathbf{r},t\right)\right|^{2},\label{fd-dens}
\end{equation}
where $f_{\beta}$ is the Fermi-Dirac occupation function depending
on the temperature $1/\beta$ and chemical potential $\mu$ at time
$t=0$ (the factor of 2 is due to spin degeneracy). Simultaneously,
the states $\phi_{j}\left({\bf r},t\right)$ evolve in time according
to the time-dependent KS equation 
\begin{equation}
i\frac{\partial\phi_{j}\left({\bf r},t\right)}{\partial t}=\left(H\left[n\left(t\right)\right]+\delta v\left({\bf r},t\right)\right)\phi_{j}\left({\bf r},t\right),\label{tdprop}
\end{equation}
where the Hamiltonian changes with time due to its implicit dependence
on the time-evolving density. A general exchange-correlation potential
would include memory effects and would therefore be non-local in time.
But here we resort to adiabatic local density approximation (ALDA),
in which $v_{xc}(n\left(\mathbf{r},t\right)$ is a function of $n\left({\bf r},t\right)$
only.

The LDA and ALDA calculations are performed using Troullier-Martins
pseudopotentials on a real-space grid with $N_{g}$ points and a spacing
of 0.6~$a_{0}$, sufficient to converge the occupied eigenvalues
to within 10~meV. The real-time propagation in its canonical form,
i.e., propagating each individual KS according to Eq.~(\ref{tdprop})
is numerically demanding because of the quadratic scaling involved,
namely $\mathcal{{\rm O}}(NN_{g})$ with a large prefactor where $N$
is the number of occupied states. In addition, there is of course
the cost of obtaining the ground state. We used deterministic DFT
which generally scales, depending on the method, as ${\rm O({\it N^{2}{\rm )-{\rm O({\it N^{3}{\rm )}}}}}}$;
the DFT was more expensive that the stochastic TDDFT method. An alternative
to the usual DFT would have been stochastic DFT,\cite{sDFT} which
would have been faster for the largest clusters.

To lower the TDDFT cost, we recently developed a stochastic framework
for TDDFT with (sub)linear scaling\cite{dan_jcp}. Instead of using
a set of all $N$ eigenstates $\left\{ \phi_{j}\right\} $, the occupied
subspace is represented by $\left| \zeta\right\rangle$ obtained as random linear
combination: 
\begin{equation}
\left|\zeta\right\rangle=\sum_{j}^{N}e^{i\theta_{j}}\sqrt{f_{\beta}\left(\varepsilon_{j},\mu\right)}\left|\phi_{j}\right\rangle\label{stochastic project}
\end{equation}
where $j$ is a state index and $\theta_{j}\in[0,2\pi]$ is a random
phase. Each $\zeta$ is a stochastic vector created using a distinct
set of random phases $\left\{ \theta_{j}\right\} $. All required
quantities are expressed using a stochastic average over $N_{\zeta}$
vectors denoted $\left\{ \cdots\right\} _{\zeta}$. For instance,
the charge density is 
\begin{equation}
n^{s}\left({\bf r}\right)=\left\{ \left|\zeta\left({\bf r}\right)\right|^{2}\right\} _{\zeta}.
\end{equation}
Since the Hamiltonian in Eq.~(\ref{KSHamiltonian}) is a functional
of the density, it also has a stochastic representation denoted $H^{s}$.

Finally, the stochastic orbitals $\left|\zeta\right\rangle$ are propagated using
a Trotter decomposition corresponding to the adiabatic stochastic
time-dependent KS equations: 
\begin{equation}
i\frac{\partial\zeta(\mathbf{r},t)}{\partial t}=\left(H^{s}\left[n\left(t\right)\right]+\delta v({\bf r},t)\right)\zeta(\mathbf{r},t),\label{stochastic time-dependent}
\end{equation}
After each propagation step $\delta\tau$, the charge density is evaluated
by 
\begin{equation}
n^{s}\left({\bf r},t\right)=\left\{ \left|\zeta\left({\bf r},t\right)\right|^{2}\right\} _{\zeta},\label{nst}
\end{equation}
and the induced dipole $\mu_{i}(t)$ is calculated from Eq.~(\ref{dipole_correlation}).
The stochastic charge density is then used to construct $H^{s}$ and
the TD procedure is repeated for the next time step $\delta\tau$,
and the propagation continues for several femtoseconds. Finally, the
absorption cross section $\sigma$ is computed from the dipole signal
$\mu_{i}\left(t\right)$ through Eq.~(\ref{dft:spectrum}).

The classical absorption spectrum is obtained from a finite-difference
time-dependent (FDTD) propagation of the Maxwell Equations, using
the MIT Electromagnetic Equation Propagation (MEEP) \cite{MEEP} open-source
package. In FDTD, the metallic nanoparticle is modeled as polarizable
material with complex dielectric permittivity given by 

\begin{equation}
\epsilon\left(\omega\right)=\epsilon^{D}\left(\omega\right)+\epsilon^{L}\left(\omega\right).\label{eq:eps-tot}
\end{equation}
Here, $\epsilon^{D}\left(\omega\right)$ is the intraband part, described
by the Drude model: 
\begin{equation}
\epsilon^{D}\left(\omega\right)=1-\frac{f_{0}\omega_{p}^{2}}{\omega\left(\omega-i\Gamma_{0}\right)},\label{eq:eps-intraband}
\end{equation}
where $\omega_{p}$ is the plasma frequency of gold and $f_{0}$,
$\Gamma_{0}$ are the intraband oscillator strength and damping constant
respectively. $\epsilon^{L}\left(\omega\right)$ is the interband
part of the dielectric permittivity, modeled as a sum of $K$ (typically
$K=2-10$) Lorentz-type terms, 
\begin{equation}
\epsilon^{L}\left(\omega\right)=\sum_{j=1}^{K}\frac{f_{j}\omega_{p}^{2}}{\left(\omega_{j}^{2}-\omega^{2}\right)+i\omega\Gamma_{j}},\label{eq:eps-interband}
\end{equation}
where $\omega_{j}$, $f_{j}$ and $\Gamma_{j}^{-1}$ are, respectively,
the j'th oscillator frequency, strength, and lifetime. We discuss
below how all these parameters were determined. 

\begin{figure*}
\includegraphics[width=0.33\textwidth]{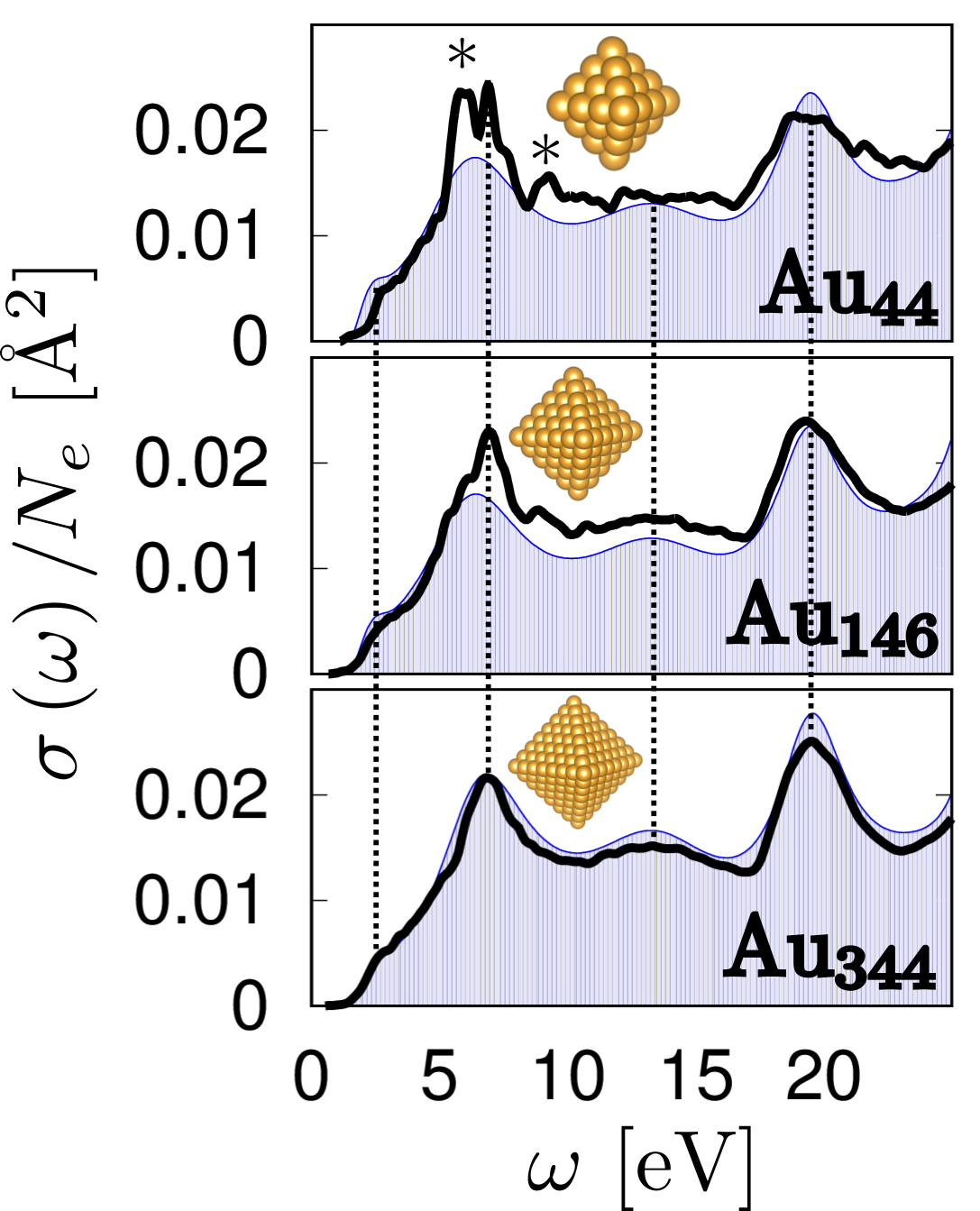}\includegraphics[width=0.33\textwidth]{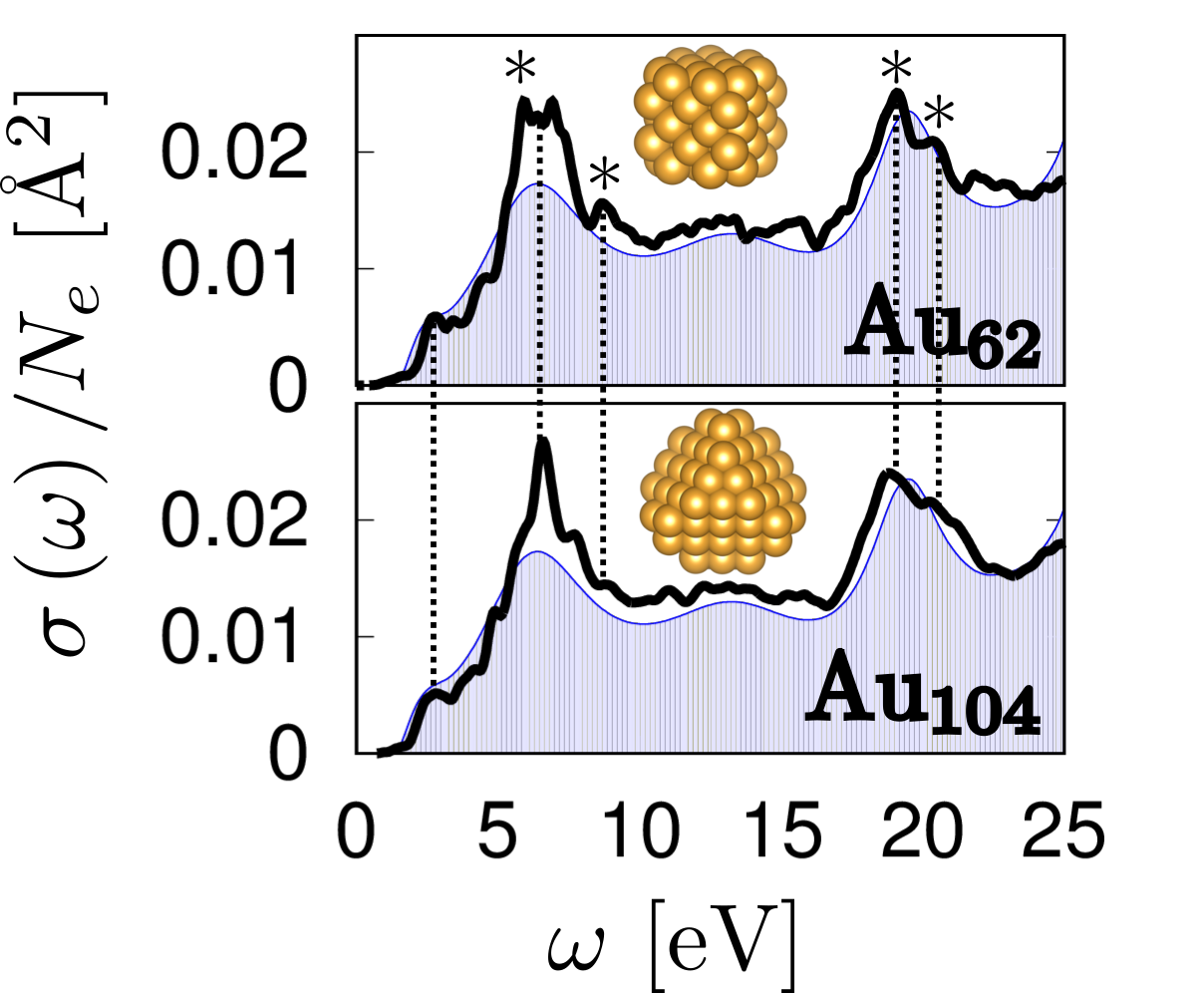}\includegraphics[width=0.33\textwidth]{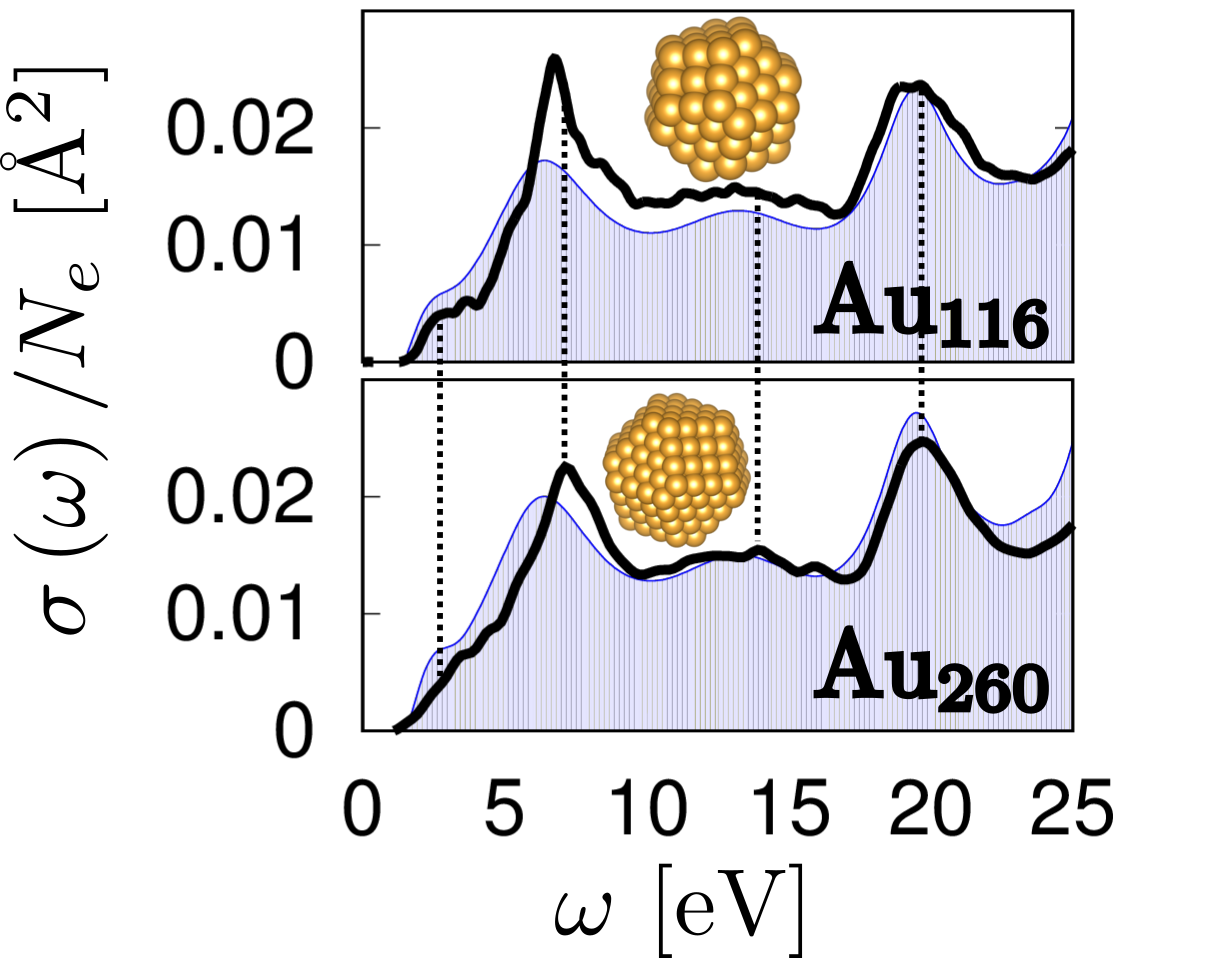}

\caption{\label{fig:spectra}Photoabsorption cross-section spectra for Au octahedra
(left panel), cubes (middle panel) and cuboctahedra (right panel).
The TDDFT (black curve) and FDTD (blue-shaded are) results are superimposed.
Vertical dashed lines show the position of the main peaks. The splittings
of the main peaks in the low (5\textendash 10~eV) and medium (17\textendash 22~eV)
frequency regions are indicated by stars ({*}). Note that the statistical
errors in the stochastic TDDFT (employed for cluster sizes bigger
than Au$_{62}$) are smaller than the line width so that the features
in the small-cluster spectra are not artifacts. }
\end{figure*}

\section*{TDDFT absorption spectra}

We consider a set of nanoparticles with octahedral, truncated cube
and cuboctahedral shapes with diameters ranging between 1.34 and 3.12
nm containing up to 3784 valence electrons. We use several closed
shell systems that were previously determined to have stable shapes
and stoichiometries, \cite{barnard_1,barnard_2} as summarized in
Table~\ref{tab:structures} and shown in Fig.~\ref{fig:spectra}.

The time-dependent electron density was obtained with stochastic TDDFT
for all systems except the smallest, Au$_{44}$ and Au$_{62}$, where
deterministic calculations were employed. The TDDFT time step was
$\delta\tau=0.03$~a.u. and for the stochastic calculations we used
$N_{\zeta}=400$ projected random vectors $\zeta$. This value of
$N_{\zeta}$ enables stable propagation up to $12\,\text{fs}$. The
stochastic propagation is especially stable for this metallic system
where the dipole is strongly damped.

\begin{table}
\tbl{\label{tab:structures}Summary of the structures. Oct., Tr.Cb., and
Cuboct. refers to octahedron, truncated cube, and cuboctahedrons,
respectively.}
{\begin{tabular}{lc c c c }
\toprule
nanoparticle & $N_{e}$ & Size {[}nm{]} & Shape\tabularnewline
\midrule
Au$_{44}$  & 484  & 1.34  & Oct. \tabularnewline
Au$_{146}$  & 1606 & 2.23  & Oct. \tabularnewline
Au$_{344}$  & 3784 & 3.12  & Oct. \tabularnewline
\midrule

Au$_{62}$  & 682  & 1.34  & Tr.Cb. \tabularnewline
Au$_{104}$  & 1144  & 1.64  & Tr.Cb. \tabularnewline
\midrule
Au$_{116}$ & 1276  & 1.61  & Cuboct. \tabularnewline
Au$_{147}$  & 1617 & 2.23  & Cuboct. \tabularnewline
\bottomrule
\end{tabular}}

\end{table}

The TDDFT optical absorption cross sections were calculated for each
system and are shown in Fig.~\ref{fig:spectra}. With the exception
of Au$_{260}$, all systems have their first absorption local maximum
at 2.8 eV, a 0.4eV blue shift from values measured in n-heptane \cite{Small_Au}.
The presence of a polarizable medium (the n-heptane solvent) results
in shifting the peaks to lower energies and explains, at least partially,
the discrepancy between experiment and our calculations of nanoparticles
in vacuum. While this peak is clearly found in the spectra of truncated
cubes and cubooctahedra, its signature is much weaker for octahedra. 

We discuss the spectral features in the three spectral regions, going
from the larger to the smaller systems (c.f. Fig.~\ref{fig:spectra}):
\begin{enumerate}
\item Intensity at lower frequencies (5 to 10 eV): For the two largest systems
(Au$_{344}$ and Au$_{260}$) only a single peak is observed (at 6.8
and 7.1 eV for the two systems, respectively). As the nanoparticle
diameter decreases the peak is split in three (where the side bands
are marked by stars). The small truncated cubes Au$_{62}$ and Au$_{104}$
always exhibit three maxima in this region. Note that except for the
smallest systems (Au$_{44}$ and Au$_{62}$), the central peak dominates
this spectral region.
\item The mid range (10 to 16 eV): The octahedral nanoparticles exhibit
a splitting of a broad maximum found at 13.3~eV for Au$_{344}$.
Spectra for other nanoparticle geometries show several local maxima. 
\item High frequencies (17 to 22~eV): Here, a single major peak is found,
shifting to higher frequencies as the system size increases. Splitting
is observed for truncated cubes (emphasized by stars above the corresponding
peaks in the figure). Note that cuboctahedra do not exhibit the splitting
as can be seen from comparison of Au$_{104}$ and Au$_{116}$ in the
figure.
\end{enumerate}

\section*{The transition from quantum to classical absorption}

\begin{figure}
\centering
\includegraphics[width=0.65\textwidth]{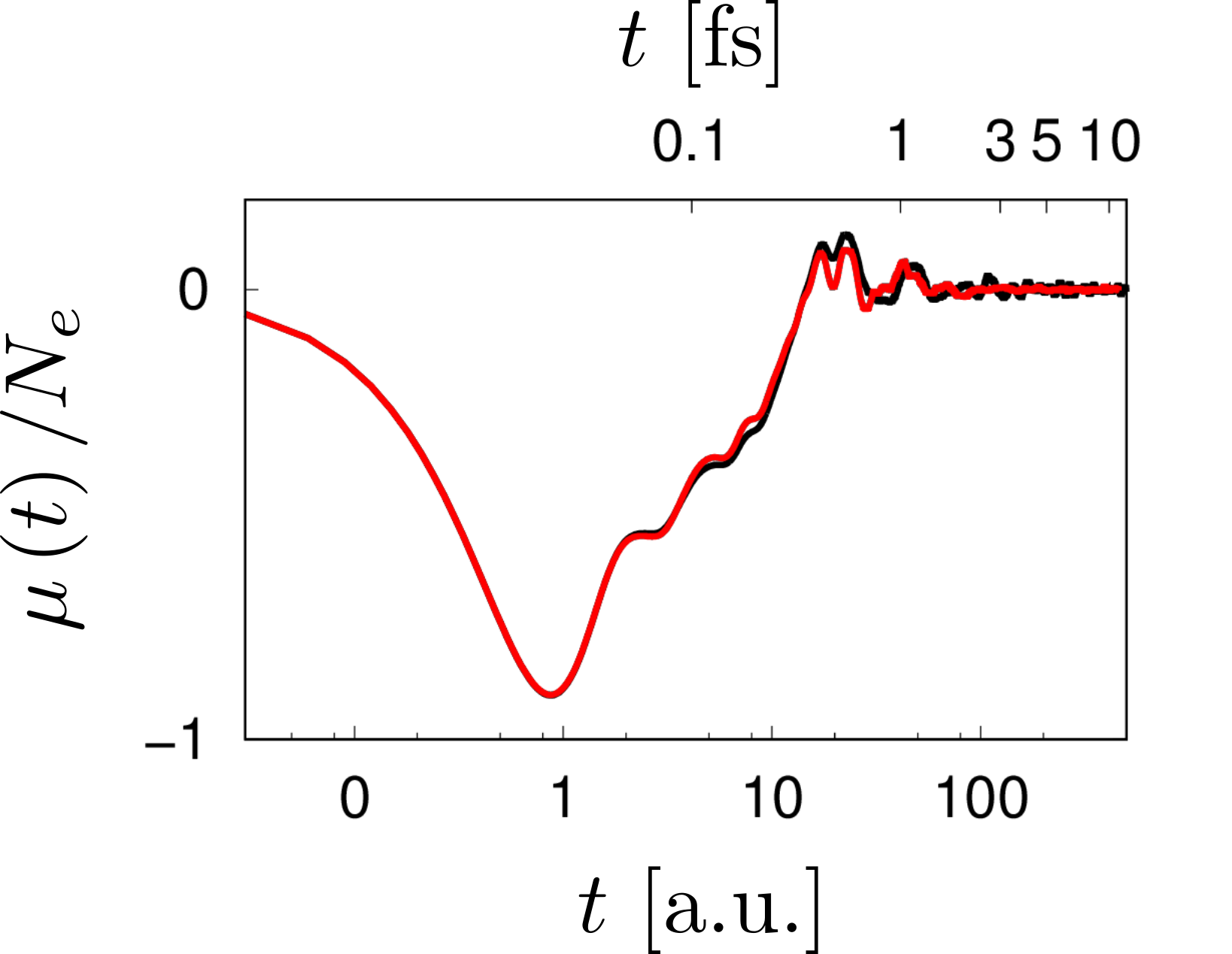} 

\caption{\label{fig:dipole}Induced dipole per valence electron as a function
of time for the smallest and the largest systems investigated.}
\end{figure}

As the system size gets smaller, the spectra change. One manifestation
is that the frequency spectrum becomes more refined, as mentioned
above. This is manifested clearly already at the level of the time-dependent
dipole moment per valence electron, $\mu_{x}\left(t\right)/N_{e}$,
(Eq.~\ref{dipole_correlation}), shown in Fig.~\ref{fig:dipole}
for the largest and smallest systems investigated. At early times,
the two $\mu_{x}\left(t\right)$ curves are almost indistinguishable
and the difference in system size shows up at later times, where the
large system's dipole decays faster. 

The next stage is to compare to the classical spectrum. For this,
we need to supply the $K$ oscillators' parameters. Typically, they
are fitted to experimentally measured real and imaginary optical dielectric
functions of the modeled bulk gold \cite{LD-1,LD_Example}. However,
our goal is to focus on the transition from quantum to classical absorption
for nanoparticles in the quantum regime, i.e., that are smaller than
3-5nm \cite{scholl2012quantum}. Therefore, we fitted the $K$ oscillators'
parameters to the TDDFT absorption cross-section for the \emph{largest
}gold nanoparticle (Au$_{344}$) and then use the same dielectric
function for all FDTD (classical) calculations. The fit used a large
frequency range, 0.5 to 25~eV, well above the 5.3eV work function
of gold \cite{WF_Au}. We used the Differential Evolution (DE) algorithm
\cite{DEA} to find oscillator parameters which minimize the following
objective function:

\begin{equation}
\chi^{2}=\int\left|\frac{\sigma_{\mathrm{TDDFT}}\left(\omega\right)-\sigma_{\mathrm{FDTD}}\left(\omega\right)}{\sigma_{\mathrm{TDDFT}}\left(\omega\right)}\right|d\omega.
\end{equation}
We found an optimal fit using $K=6$ oscillators shown in the Table~\ref{tab:FDTD-fit}. 

\begin{table}
\tbl{\label{tab:FDTD-fit}Values of the Lorentz-Drude model parameters.
Values are given in eV. The plasma frequency was taken from Ref.~\cite{LD_Example}
as 9.03 eV.}
{\begin{tabular}{l c c c c}
\toprule
Oscillator  & $\omega_{j}$  & $f_{j}$  & $\Gamma_{j}$ \tabularnewline
\midrule
0  & 0.000  & 1.234  & 0.000 \tabularnewline

1  & 0.523  & 0.000  & 1.735 \tabularnewline

2  & 4.000 & 2.479  & 5.506 \tabularnewline

3  & 12.921  & 1.980  & 7.467 \tabularnewline

4  & 18.831  & 2.405  & 3.223 \tabularnewline

5  & 25.568  & 20.000  & 3.325 \tabularnewline
\bottomrule
\end{tabular}}

\end{table}

Comparing the classical and TDDFT spectra in Fig.~\ref{fig:spectra}
reveals that for all systems smaller than Au$_{344}$ the finer observed
behavior is not captured well. Only the first peak at 2.8 eV is correctly
found to be insensitive to system size and shape, in agreement with
results from stochastic TDDFT calculations. For all systems, FDTD
spectra is very smooth and show only three major peaks.

The first maximum shows slight size dependence, its position for Au$_{344}$
and Au$_{44}$ is shifted by 0.4~eV to lower frequencies for the
smaller nanoparticle, but no splitting is observed. The position of
other peaks in the FDTD spectra remains unchanged. Furthermore, TDDFT
captures slight shape difference between Au$_{116}$ and Au$_{104}$
as symmetrical splitting of the peak at 19.7~eV, which is completely
absent in the FDTD results. We note that the lack of shape dependence
in the FDTD spectra cannot be attributed to the coarseness of the
real-space grid employed in the classical simulation since that was
well converged even for the smallest system investigated (Au$_{44}$).
Instead, we attribute the changes in the spectra to quantum signatures
and decreased DOS population in the Fermi energy. 

For small nanoparticle sizes, quantum confinement effects dominate
the spectra and individual electronic states couple more strongly
to the nanoparticle surface. Furthermore, broad spectral features
break down to individual electronic transitions leading to multiple
sharp maxima. Within the classical FDTD approach, the shape of the
system is treated as homogeneous and isotropic polarizable continuum
but its precise geometry (octahedron, cuboctahedron or truncated cube)
has only negligible effect on the resulting absorption.

\section*{Summary and Conclusions}

We used an FDTD Maxwell solver and our newly developed sTDDFT method
to investigate the effect of size and shape on the absorption cross
section of gold nanoparticles as large as 3nm. The sTDDFT calculated
spectra show features that are consistent with the classical Maxwell
theory as the system approaches the bulk limit. Moreover, the fine
structure, significant in the small systems, agree with our a-priori
expectation: in large particles where the properties are essentially
metallic, the infinite number of states will result in a continuous,
smooth absorption spectrum. As the systems become smaller, the finite
number of states lead to discretization and splittings in the spectra.
This is in line with experimental results where individual gold nanoparticles
below 3 nm show reduction in the DOS of the Fermi level \cite{DOSAu},
and is reflected by the jagged nature of the spectrum for small clusters. 

Based on these observations, we conclude that even when we force the
classical methods to fit qualitatively the TDDFT optical properties
of gold nanoparticles at 3 nm, it will lose the finer detail for smaller
nanoparticles. 

Our studies verify that stochastic TDDFT is a valid technique for
calculating absorption cross-section of bulk-like gold nanoparticles.
They also show that or particles smaller than 3 nm, fully quantum
methods are required to predict the finer details of the absorption
cross-section spectra.

\section*{Acknowledgments}

D.N. acknowledges support by the NSF grant DMR/BSF-1611382, E.R. acknowledge
support by NSF grant CHE-1465064, and R.B. Acknowledges the support
of the Binational Science Foundation grant BSF 2015687. The calculations
were performed as part of the XSEDE \cite{towns2014xsede} computational
project TG-CHE170058. 

\bibliographystyle{unsrt}
\bibliography{test}
 
\end{document}